\documentclass[conference]{IEEEtran}
\IEEEoverridecommandlockouts
\usepackage{cite}
\usepackage{amsmath,amssymb,amsfonts}
\usepackage{algorithmic}
\usepackage{graphicx}
\usepackage{textcomp}
\usepackage{xcolor}
\def\BibTeX{{\rm B\kern-.05em{\sc i\kern-.025em b}\kern-.08em
    T\kern-.1667em\lower.7ex\hbox{E}\kern-.125emX}}

\usepackage[acronym]{glossaries}
\usepackage{amsmath}
\usepackage{bm}
\usepackage{amsthm}

\ifCLASSOPTIONcompsoc
\usepackage[caption=false,font=normalsize,labelfon
t=sf,textfont=sf]{subfig}
\else
\usepackage[caption=false,font=footnotesize]{subfi
g}
\fi

\newacronym{aoi}{AoI}{Age of Information}
\newacronym{fcfs}{FCFS}{first-come-first-served}
\newacronym{lcfs}{LCFS}{last-come-first-served}
\newacronym{arq}{ARQ}{automatic repeat-request}
\newacronym{harq}{HARQ}{hybrid automatic repeat-request}
\newacronym{mdp}{MDP}{Markov decision process}
\newacronym{mac}{MAC}{multiple access channel}
\newacronym{tdma}{TDMA}{time devision multiple access}
\newacronym{mc}{MC}{multi-connectivity}
\newacronym{pd}{PD}{packet duplication}
\newacronym{mp}{MP}{multiplexing}
\newacronym{cs}{CS}{codeword splitting}
\newacronym{ms}{MS}{message splitting}
\newacronym{sc}{SC}{single channel}
\newacronym{awgn}{AWGN}{additive white Gaussian noise}
\newacronym{snr}{SNR}{signal to noise ratio}
\newacronym{rat}{RAT}{radio access technology}
\newacronym{psd}{PSD}{positive semidefiniteness}

\newtheorem{theorem}{Theorem}
\newtheorem{corollary}{Corollary}
\newtheorem{remark}{Remark}

\begin{document}

\title{Age of Information in Short Packet Multi-Connectivity Links\\
\thanks{The presented research is based on the joint research project “RePro”, which is funded
by the Federal Ministry of Education and Research (BMBF, grant number
16KISR015K). The work of M. Mross and E. Jorswieck is supported in part by the Federal Ministry of Education and Research of Germany in the program of "Souverän. Digital. Vernetzt." Joint project 6G-RIC, project identification number: 16KISK031. The authors are responsible for the content of this publication.}
}

\author{\IEEEauthorblockN{Mojan Wegener, Marcel A. Mross, Eduard A. Jorswieck}
\IEEEauthorblockA{
Institute for Communications Technology, Technische Universität Braunschweig, Germany \\
\{mojan.wegener, m.mross, e.jorswieck\}@tu-braunschweig.de
} \\
}

\maketitle

\begin{abstract}
     In this paper, we investigate \gls{mc} schemes in the context of status update systems with short payloads. As the performance metric, we use the \gls{aoi}. Due to short payloads, transmission errors must be taken into account.
    In addition to the well-known schemes of packet duplication, message splitting, and multiplexing, we propose a codeword splitting scheme, where each status update is jointly encoded across multiple channels. We derive closed-form expressions for the average \gls{aoi} for the different schemes and optimize their corresponding parameters. We show that the \gls{aoi} of the multiplexing scheme is a convex function in the offset parameters and use this result to prove that for homogeneous channels, the multiplexing scheme outperforms the packet duplication scheme, which is outperformed by the codeword splitting scheme. Analytical comparisons and numerical evaluations show that the codeword splitting scheme achieves the lowest average \gls{aoi} when joint coding is feasible.
    In scenarios where joint coding is not feasible, whether message splitting or multiplexing results in a lower average \gls{aoi} depends on the specific parameters.
\end{abstract}

\begin{IEEEkeywords}
    Age of Information, Multi-Connectivity, Short Packet Communication, Timely Status Updates
\end{IEEEkeywords}

\section{Introduction}

Timely status updates play a crucial role in future communication systems. In many use cases, such as remote-controlled vehicles and factory automation, the receiver is primarily interested in the most recent available status update of a given source. To quantify the timeliness of status updates, the \gls{aoi} metric was introduced in \cite{kaul_real-time_2012}. This metric measures the time elapsed since the generation of the last successfully received status update. Unlike related metrics such as delay and inter-packet gap, the \gls{aoi} metric captures both aspects simultaneously.  

Since its introduction, \gls{aoi} has been widely studied in the research community. In \cite{kaul_real-time_2012, kam_effect_2016, yates_status_2018, yates_age_2019}, the authors analyze \gls{aoi} primarily from a queuing perspective. Specifically, in \cite{kaul_real-time_2012}, the \gls{aoi} for M/M/1 and M/D/1 queues under the \gls{fcfs} policy is examined. In \cite{kam_effect_2016}, the effect of multiple parallel servers is incorporated, and the time-average \gls{aoi} for M/M/2 and M/M/$\infty$ queuing systems under the \gls{fcfs} policy is studied. The work in \cite{yates_status_2018} extends this analysis to M/M/c systems under the last-come-first-served policy, using a stochastic hybrid system approach. In \cite{yates_age_2019}, the same approach is used to derive the \gls{aoi} for a system with multiple sources, competing for a single server. 
%The authors in \cite{chen_timeliness_2024} analyze a system where sources with random status update arrivals and zero transmission times are used in parallel with generate-at-will sources with random transmission times using the stochastic hybrid systems approach.
In all these studies, transmission times are modeled as i.i.d. random variables, and it is assumed that no transmission errors occur. 

Another line of research \cite{yu_average_2020, wang_age_2019, sung_age_2022} investigates the trade-off between transmission time, determined by the number of error correction symbols used to transmit status updates, and the probability of decoding error. The key tool in this line of research is the second-order normal approximation from \cite{polyanskiy_channel_2010}, which relates the error probability to the codeword length. In \cite{yu_average_2020}, the time-average \gls{aoi} for an \gls{arq} protocol is compared with a simple protocol that does not use feedback messages, and their optimal codeword lengths are determined. The study in \cite{wang_age_2019} focuses on comparing different packet management schemes, such as preemption, non-preemption, and retransmission, in the context of \gls{aoi}. The authors in \cite{sung_age_2022} analyze the \gls{aoi} for short blocklengths in a downlink cellular network scenario with fading. 

Another important topic in the field of wireless communication systems is \gls{mc}. The three fundamental \gls{mc} transmission schemes are \gls{pd}, \gls{mp}, and \gls{ms}, which have been extensively analyzed and compared in prior research \cite{suer_multi-connectivity_2020}. While most studies focus on latency and throughput, little work investigates the \gls{mc} transmission schemes in the context of \gls{aoi} in the finite blocklength regime. In \cite{chiariotti_peak_2021, chiariotti_latency_2022}, the peak \gls{aoi} is analyzed for \gls{mc} schemes where status updates are jointly encoded over multiple links using an erasure code. 
Most closely related to our work is \cite{cao_multiplexing_2023}, which compares multiplexing and diversity transmission in terms of the average \gls{aoi} based on a Markov model. The study assumes Poisson arrivals, meaning the diversity–multiplexing trade-off is primarily determined by the mean arrival rate. In contrast, we consider a generate-at-will source and take the optimization of the parameters of each scheme into account, leading to a different approach to comparing the \gls{mc} schemes.
%In \cite{cao_multiplexing_2023}, multiplexing and diversity transmission are compared in terms of average \gls{aoi} for a system with Poisson status update arrivals. 

\subsection{Our Contributions}

In this work, we focus on the average \gls{aoi} of a system where a generate-at-will source transmits status updates over $N$ parallel \gls{awgn} channels. We consider a regime with short payloads, in which transmission errors must be accounted for. Our key contributions are as follows:  

\begin{itemize}
    \item We derive closed-form expressions for the average \gls{aoi} of the basic transmission schemes: packet duplication, multiplexing, and message splitting. Additionally, we propose a \gls{cs} scheme and derive the corresponding \gls{aoi}. 
    \item Using the closed-form expressions, we derive the optimal schedule for the \gls{mp} scheme and prove that for homogeneous channels, the \gls{mp} scheme outperforms the \gls{pd} scheme, while the \gls{cs} scheme outperforms the \gls{mp} scheme. 
    \item Based on our analytical results, we perform numerical evaluations of the different \gls{mc} schemes. The results show that the \gls{cs} scheme outperforms other schemes when joint coding is feasible. When joint coding is not feasible, the choice between message splitting and multiplexing depends on the system parameters. 
    For heterogeneous channels, we show how to optimize the message splits for the \gls{ms} scheme. 
\end{itemize}

\section{System Model}\label{sec:Systemmodel}

\subsection{Communication System}

\begin{figure}[t]
\centerline{\includegraphics[width=\linewidth]{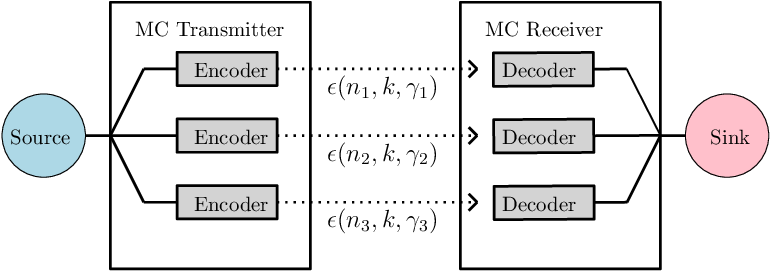}}
\caption{Schematic visualization of the system model with $N=3$ channels.}
\label{fig:SystemModel}
\end{figure}

We consider a single source-sink pair, which is connected by $N$ wireless channels. Each channel is modeled as an \gls{awgn}-channel with equal bandwidth and a \gls{snr} of $\gamma_i$ which can vary across the channel index $i \in [0,1,..N-1]$. The source is modeled as a generate-at-will source. Each status update consists of $k$ bits. We consider a system where the status updates are relatively small, so that transmission errors at short blocklengths have to be taken into account. We therefore model the transmission error probability of a single channel by the second-order normal approximation from \cite{polyanskiy_channel_2010}, which is given by
\begin{align}\label{eq:BasicEpsilonExpression}
    	\epsilon(n,k,\gamma) \approx Q \left(  \frac{\frac{1}{2}\log_2(1+\gamma)-\frac{k}{n}}{\log_2(e)\sqrt{\frac{1}{2n}(1-\frac{1}{(1+\gamma)^2})}} \right),
\end{align}
where $n$ denotes the blocklength, $k$ denotes the number of message bits and $\gamma$ is the \gls{snr} of the channel. We assume that the error events are distributed i.i.d. across channels and time. 
The system model is visualized in Fig.\ref{fig:SystemModel}. 
\subsection{The Age of Information Metric}
In this subsection, we introduce the \gls{aoi}-metric and describe the evolution of the metric in the context of transmission errors. Let $r(t)$ be the generation time of the last successfully decoded status update at the receiver side. The \gls{aoi} $\Delta(t)$ is the random process defined by
\begin{align}
    \Delta(t) = t - r(t).
\end{align}
Therefore, at the points in time where new status updates are successfully decoded at the receiver, the \gls{aoi} is equal to the delay of these updates, and in between these points in time, the \gls{aoi} increases with unit slope. The primary performance metric analyzed in this paper is the time average \gls{aoi} $\Bar{\Delta}$, which is defined as
\begin{align}
    \bar{\Delta} = \lim_{T \to \infty} \frac{1}{T} \int_{0}^{T} \Delta(t) dt .
\end{align}
In this paper all time intervals are given as multiples of the transmission time $T_S$ of one symbol, which depends on the used bandwidth. 

\subsection{System Operation}

In the system model, no feedback about successful transmissions is available at the transmitter. We therefore investigate the following simple periodic transmission schemes for the \gls{mc} scenario:

\begin{itemize}
    \item \textbf{SC}: In the \gls{sc}  case, the blocklength $n$ is fixed. Therefore, every $n$ symbols a new status update is encoded and transmitted over the channel. 
    \item \textbf{PD}: For the \gls{pd} scheme, each status update is encoded, and the encoded update is transmitted over all $N$ channels in parallel using the same fixed blocklength $n$. 
    \item \textbf{MP}: In the \gls{mp} scheme, each of the $N$ channels operates similarly to the SC case with a common blocklength $n$, but with potentially different offsets between the periodic transmission schedules of the single channels.
    \item \textbf{CS}: In the \gls{cs} scheme, each status update is encoded over all channels simultaneously. Thus, a single codeword is split into $N$ equal fragments, which are then transmitted in parallel over the $N$ channels. 
    \item \textbf{MS}: In the \gls{ms} scheme, the $k$ message bits of one status update are split into $N$ fragments $k_0, k_1,..,k_{N-1}$ with $\sum_{i=0}^{N-1} k_i = k$, the fragments are distributed over the $N$ channels and each fragment is encoded with a fixed blocklength $n$. 
\end{itemize}

\begin{figure}[ht]
\centerline{\includegraphics[width=0.9\linewidth]{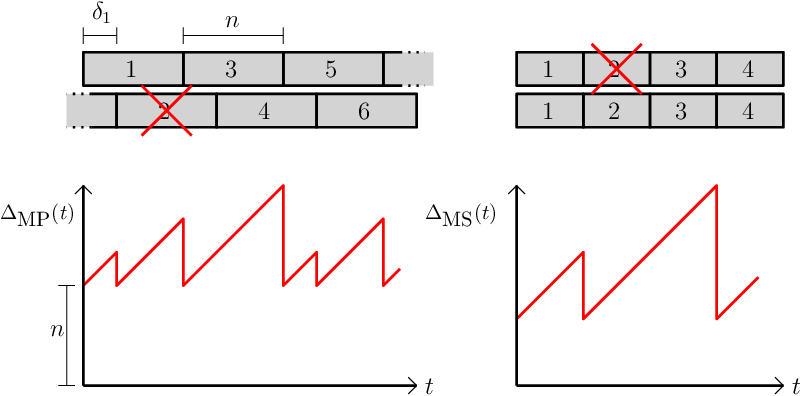}}
\caption{Schematic Comparison of the multiplexing scheme and the message splitting scheme, with $N=2$ channels. Channel 0 is always used as the reference for shift parameters $\delta_i$, so only $\delta_1$ is shown on the left since $\delta_0 = 0$. }
\label{fig:MultiplexingSchematic}
\end{figure}

The system operation of the \gls{mp} scheme and the \gls{ms} scheme are visualized in Fig. \ref{fig:MultiplexingSchematic}. 
In the next section, we derive expressions for the average \gls{aoi} of the different schemes.

\section{Average AoI Analysis}

In the first part of this section, closed-form expressions for the average \gls{aoi} for the different transmission schemes are derived. In the second part, we discuss the performance of the schemes analytically and by numerical evaluations.\footnote{For clarity and notational simplicity, we treat variables as continuous in the analytical derivations. The integer constraints are fully taken into account in the numerical evaluations.}

\subsection{Analytical AoI Analysis}\label{subsec:AnalyticalAoIAnalysis}

For the schemes SC, PD, MS, and CS, the \gls{aoi} process has the form of a renewal reward process. Every renewal epoch begins with a successful decoding at the receiver, with the reward being the \gls{aoi} integrated over the length of the epoch. By the basic renewal reward theorem \cite[Theorem 5.4.5]{gallager_stochastic_2013} and by noting that the area under the \gls{aoi} curve during one epoch is given by a rectangle and a triangle, the average \gls{aoi} can be expressed as

\begin{align}
    \bar{\Delta} = \frac{\mathbb{E}[\frac{1}{2}Y^2 + n Y] }{\mathbb{E}[Y]} = \frac{\mathbb{E}[Y^2]}{2\mathbb{E}[Y]} + n,\label{eq:generalAoIexpression}
\end{align}
where $Y$ is the time between two successfully decoded status updates and $n$ is the blocklength used.
Using the recursive approach from \cite{wang_age_2019}, the first two moments of the random variable $Y$ can be calculated as
\begin{align}
    \mathbb{E}(Y) & = (1-\epsilon) n + \epsilon \mathbb{E}[n + Y] \label{eq:firstMomentSC} \\
    \mathbb{E}[ Y^2 ] & =  (1-\epsilon) n^2 + \epsilon \mathbb{E}[(n + Y)^2]  \label{eq:secondMomentSC} 
\end{align}
where $\epsilon$ denotes the error probability, which depends on the transmission scheme: 

\textbf{SC scheme}: $\epsilon_{\text{SC}}(n)$ is given by $\epsilon(n, k, \gamma$).

\textbf{PD scheme}: An error occurs only if all the parallel transmissions fail. Therefore the error probability is given by $\epsilon_{\text{PD}}(n) = \prod_{i=0}^{N-1} \epsilon(n,k, \gamma_i)$. 

\textbf{MS scheme}: An error occurs if at least one of the $N$ parallel transmissions fails. Therefore the error probability can be written as  $\epsilon_{\text{MS}}(n) = 1- \prod_{i=0}^{N-1} (1-\epsilon(n,k_i, \gamma_i))$. 

\textbf{CS scheme}: Here \cite[Theorem 78]{polyanskiy_channel_2010-1} can be used. This theorem gives the second-order rate normal approximation for the transmission over $N$ parallel \gls{awgn} channels. Note that while in the theorem there is only a single power constraint over all $N$ channels and an optimization over the power allocation is done, the theorem is also valid for the case of individual power constraints for the individual channels. Neglecting the higher-order terms, as in the single-channel case,  $\epsilon_{\text{CS}}$ can be expressed as
    \begin{align}\label{eq:epsilonSpl}
        \epsilon_{\text{CS}}(n) =  Q\left(\frac{\sum_{i=1}^{N} \frac{1}{2}\log_2(1+\gamma_i) - \frac{k}{n}}{\log_2(e)\sqrt{\frac{1}{2n}\sum_{i=1}^{N} \left(1-\frac{1}{(1+\gamma_i)^2}\right)}} \right).
    \end{align}
Solving (\ref{eq:firstMomentSC}) for $\mathbb{E}[Y]$ and (\ref{eq:secondMomentSC}) for $\mathbb{E}[Y^2]$ and combining everything together in (\ref{eq:generalAoIexpression}) gives the following expression for average \gls{aoi}:

\begin{align}
    \bar{\Delta}_{\chi}(n) = \frac{n(1+\epsilon_{\chi}(n))}{2(1-\epsilon_{\chi}(n))} + n,
\end{align}
where $\chi \in \{\text{SC}, \text{MS}, \text{PD}, \text{CS} \}$ denotes the used \gls{mc} scheme. 
For all the transmission schemes described above, an optimization over the blocklength $n$ has to be done, and for the MS scheme, an additional optimization over the message splits $k_0, k_1,..,k_{N-1}$ has to be done. 
With the expressions for the error probability, we define the minimal achievable average \gls{aoi} of the schemes \gls{sc}, \gls{pd} and \gls{cs} as
\begin{align}
    \bar{\Delta}^*_{\chi} = \min_{n > 0}  \bar{\Delta}_{\chi}(n) \quad \!\!\!\!\text{for $\chi \in \{ \text{SC}, \text{PD}, \text{CS} \}$},
\end{align}
and for \gls{ms} as
\begin{align}
    \bar{\Delta}^*_{\text{MS}} & = \!\!\!\!\!\!\! \min_{n > 0, k_0,..,k_{N-1}} 
 \!\!\!\!\!\! \bar{\Delta}_{\text{MS}}(n) \label{eq:messageSplitOptimizationProblem} \\
     \text{subject to } & k_i \geq 0, \qquad
     \sum_{i=0}^{N-1}k_i = k. 
\end{align}
Deriving analytical results for the \gls{mp} scheme becomes highly challenging when considering parallel channels with different \gls{snr} values. Therefore, we focus on the simplified case of homogeneous channels, where all channels have the same \gls{snr}, i.e., $\gamma_0 = \gamma_1 = .. = \gamma_{N-1}$.
Note that while all channels use the same blocklength $n$, they may have different offsets. We define the shift parameters $\delta_0, \delta_1, .. ,\delta_{N-1}$ relative to the first channel i.e. $\delta_0 = 0$. The shift parameters are visualized in Fig. \ref{fig:MultiplexingSchematic} on the left. 
Regarding the optimization of the shift parameters, we have the following theorem:
\begin{theorem}\label{theo:multiplexingConvexity}
    Let $\bar{\Delta}_{\text{MP}}(n,\delta_0, \delta_1,..,\delta_{N-1})$ denote the average \gls{aoi} for the \gls{mp} scheme, which is a function of the blocklength $n$ and the shift parameters $\delta_0,..,\delta_{N-1}$. If all channels have the same \gls{snr} i.e. $\gamma_0 = \gamma_1 = .. = \gamma_{N-1}$, the following holds:
    \begin{itemize}
        \item The function $\bar{\Delta}_{\text{MP}}(n,\delta_0, \delta_1,..,\delta_{N-1})$ is convex in the shift parameters $\delta_0, \delta_1,..,\delta_{N-1}$. 
        \item For a fixed $n$ the function attains its minimum at
        \begin{align}
            \delta_i = i \cdot \frac{n}{N} \quad \text{for $i \in [0, N-1]$},
        \end{align}
        and the minimum is given by
        \begin{align}
            \min_{\delta_0, \delta_1,..,\delta_{N-1}} \!\!\!\!\!\! \bar{\Delta}_{\text{MP}}(n,\delta_0,..,\delta_{N-1}) \! = \! \frac{n(1+\epsilon_{\text{SC}}(n))}{2N(1-\epsilon_{\text{SC}}(n))}\!  + \! n.
        \end{align}
    \end{itemize}
\end{theorem} 
For the proof, refer to Appendix A
Based on Theorem \ref{theo:multiplexingConvexity}, the minimum average \gls{aoi} of the \gls{mp} scheme for the case of $N$ channels with equal \gls{snr} is given by:
\begin{align}
    \bar{\Delta}^*_{\text{MP}} = \min_{n > 0} \frac{n(1+\epsilon_{\text{SC}}(n))}{2N(1-\epsilon_{\text{SC}}(n))} + n.
\end{align}
\subsection{Comparison of the MC schemes}\label{sec:ComparisonEqualSNR}
\begin{figure*}[!t]
\centering
\subfloat[$k=16$]{\includegraphics[width=3in]{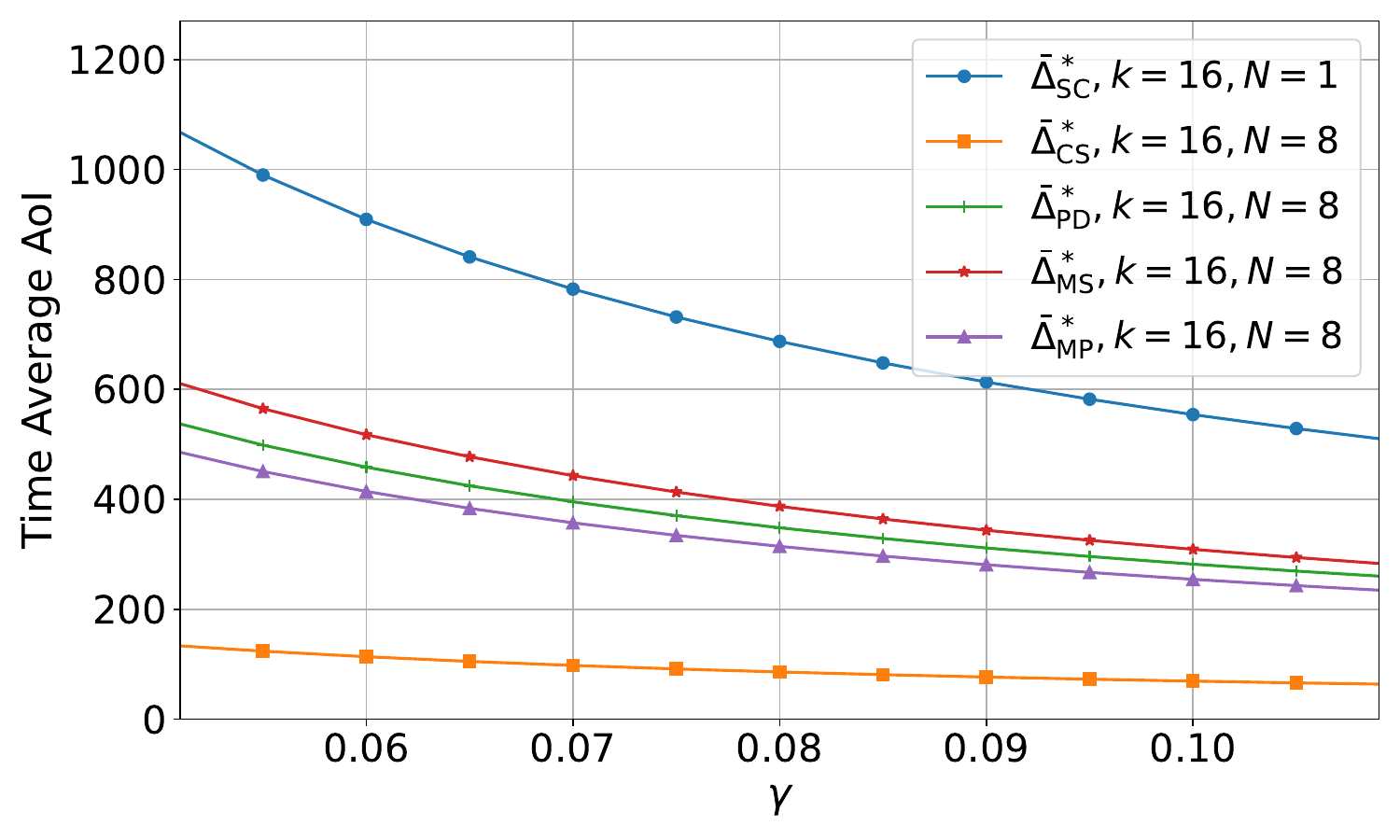}
\label{fig_first_case}}
\hfil
\subfloat[$k=32$]{\includegraphics[width=3in]{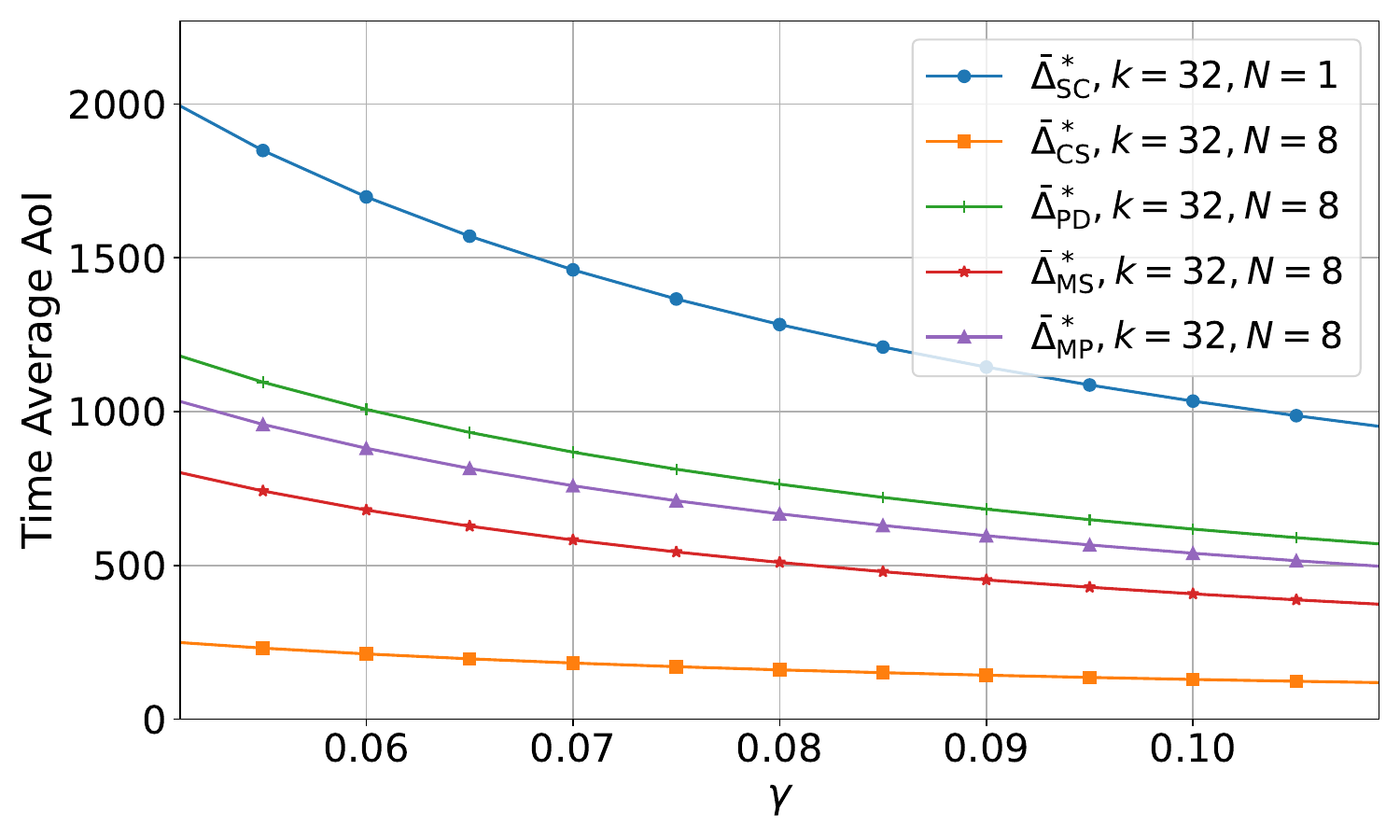}
\label{fig_second_case}}
\caption{The mininal average AoI achievable by different MC schemes for $N$ channels with equal \gls{snr} vs the \gls{snr} $\gamma$.}
\label{fig:AoIEqualChannelComparison}
\end{figure*}

In this section, we discuss the different \gls{mc} schemes based on a comparison of the analytical expressions and numerical evaluations for the case of $N$ parallel channels with equal \gls{snr}. We begin by comparing the \gls{pd} scheme and the \gls{mp} scheme. Note that, in the case of $N$ channels with equal \gls{snr}, the \gls{pd} scheme can be viewed as a special case of the \gls{mp} scheme, where the shift parameters are chosen as $\delta_0 = \delta_1 = \ldots = \delta_{N-1} = 0$. By Theorem \ref{theo:multiplexingConvexity}, we know that the average \gls{aoi} with these shift parameters is greater than or equal to the average \gls{aoi} achieved by the \gls{mp} scheme with the optimal shift parameters $\delta_i = i \cdot \frac{n}{N}$. This is valid for all $n$. Therefore, we conclude that $ \bar{\Delta}^*_{\text{MP}} \leq  \bar{\Delta}^*_{\text{PD}}$.
Comparing the \gls{mp} scheme and the \gls{cs} scheme for a fixed blocklength $n$ is challenging due to the complexity of (\ref{eq:BasicEpsilonExpression}). To address this, we use the following approach: rather than expressing the average \gls{aoi} in terms of the blocklength $n$, as done in Section \ref{subsec:AnalyticalAoIAnalysis}, we express it in terms of the actual codeword length $n'$. Note that we have the relation $n' = N\cdot n$ for the \gls{cs} scheme and $n' = n$ for the \gls{mp} scheme. With this, the difference  $d_{\text{MP,CS}}(n')$ between the average \gls{aoi} of the two schemes can be expressed as 
\begin{align}
    & d_{\text{MP,CS}}(n',N)  = \nonumber \\
    & \frac{n'(1+\epsilon_{\text{SC}}(n')}{2N(1-\epsilon_{\text{SC}}(n'))} + n' - \left( \frac{n'(1+\epsilon_{\text{SC}}(n')}{2N(1-\epsilon_{\text{SC}}(n'))} + \frac{n'}{N} \right) \\
    & =  \frac{(N-1)n'}{N}
\end{align}
which is always positive for $N \geq 1$ and $n' \geq 0$.
Combining this with the previous result, we establish the following theorem:
\begin{theorem}
Let $\bar{\Delta}^*_{\text{CS}}$, $\bar{\Delta}^*_{\text{MP}}$, $\bar{\Delta}^*_{\text{PD}}$ denote the minimal average \gls{aoi} for the \gls{cs}, \gls{mp} and \gls{pd} scheme for $N
$ channels with equal SNR. Then it holds that   
\begin{align}
    \bar{\Delta}^*_{\text{CS}} \leq  \bar{\Delta}^*_{\text{MP}} \leq  \bar{\Delta}^*_{\text{PD}}  
\end{align}
\end{theorem}
The comparison thus far does not include the \gls{ms} scheme.  
A direct analytical comparison of this scheme with the other schemes is challenging due to the complexity of (\ref{eq:BasicEpsilonExpression}). 
Therefore, we perform numerical evaluations based on the closed-form expressions for the average \gls{aoi} of the different \gls{mc} schemes. 
For the PD, MP, SC, and CS schemes, we optimize $n$. 
For the MS scheme, we optimize $n$ as well as the possible message splits $k_0, k_1, \dots, k_{N-1}$ through an exhaustive search. 
The results are shown in Fig. \ref{fig:AoIEqualChannelComparison} for two different values of the message size $k$. 
The figures indicate that the relative performance of the MS scheme compared to the MP scheme and the PD scheme depends on the specific parameters $k$ and $\gamma$.  
Based on extensive numerical evaluations, we find that the MS scheme achieves a lower average \gls{aoi} across a significant portion of the parameter space. 
However, the MP scheme outperforms the MS scheme in regions with a small number of message bits $k$ and a high number of parallel channels. 
The relatively poor performance of the MS scheme in scenarios with a high number of parallel channels may be attributed to the requirement that all message fragments must be successfully decoded. 
For large $N$, this effect cannot be fully compensated by the reduction in error probability per individual channel, which results from the corresponding decrease in transmission rate.
\section{Optimizing the Message Split}
In the previous section, we observed that for a large portion of the parameter space, the MS scheme outperforms the PD scheme and the MP scheme. At the same time, the MS scheme is outperformed by the CS scheme under the condition that $\epsilon_{\text{CS}}(n) \leq \epsilon_{\text{MS}}(n)$, which holds when maximum likelihood decoding is used. Note that the expressions for $\epsilon_{\text{CS}}(n)$ and $\epsilon_{\text{MS}}(n)$ are based on an approximation of a suboptimal decoder. Thus, it may be possible that formally the expressions do not fulfill the inequality, but this could then be attributed to the sub-optimality of the second-order rate normal approximation from \cite{polyanskiy_channel_2010}. A disadvantage of the CS scheme is that it requires joint encoding and decoding across multiple channels. This may not always be possible, for example, in a multi-\gls{rat} scenario. Therefore, in this section, we focus on solving the optimization problem (\ref{eq:messageSplitOptimizationProblem}) for the scenario of $N$ channels with possible different \gls{snr}-values.
Note that $\bar{\Delta}_{\text{MS}}$ is a strictly increasing function of the error probability $\epsilon_{\text{MS}}(n)$. Thus, for a fixed $n$, the probability of a successful transmission must be maximized. This leads to the following optimization problem:
\begin{align}
    & \max_{k_0, k_1, ..,k_{N-1}} \prod_{i=0}^{N-1} (1-\epsilon(n,k_i, \gamma_i)) \\
    & \text{subject to } \sum_{i=0}^{N-1} k_i = k, \quad
    k_i \geq 0, \text{ for $i \in [0, N-1]$}.
\end{align}
Applying the logarithm to the objective function and noting that the argument of the Q-function in (\ref{eq:BasicEpsilonExpression}) is an affine function in $k$, we obtain the following equivalent optimization problem:
\begin{align}
    & \max_{k_0, k_1, ..,k_{N-1}} \sum_{i=0}^{N-1} \log \left( \Phi\left( A_i\left( B_i - \frac{k_i}{n} \right) \label{eq:MessageSplitTransformedOptimization} \right) \right) \\
    & \text{subject to } \sum_{i=0}^{N-1} k_i = k, \quad
    k_i \geq 0, \text{ for $i \in [0, N-1]$},
\end{align}
where $A_i = \frac{1}{\log_2(e)\sqrt{\frac{1}{2n}(1-\frac{1}{(1+\gamma_i)^2})}}$, $B_i = \frac{1}{2}\log_2(1+\gamma_i)$ and $\Phi(x) = \int_{-\infty}^{x}\frac{1}{\sqrt{2\pi}} \exp(\frac{-t^2}{2})dt$ denotes the normal CDF. It is well known that \( \log(\Phi(x)) \) is concave \cite{boydConvexOpt2006}. Since the argument of \( \Phi(x) \) is an affine function and the constraints are linear, the optimization problem in \eqref{eq:MessageSplitTransformedOptimization} is convex and can be efficiently solved using numerical solvers such as CVX \cite{diamond2016cvxpy}. The joint optimization of $n$ and $\bm{k} = (k_0,..,k_{N-1})^T$ can be performed by numerically solving (\ref{eq:MessageSplitTransformedOptimization}) for each value of $n$ and selecting the minimum of the corresponding average \gls{aoi} values.  
In the special case where all \( N \) channels have equal \gls{snr}, the objective function is symmetric, meaning that swapping any two variables does not change its value. Due to this symmetry and concavity, the function is Schur-concave \cite{jorswieck_majorization_2006}. For a Schur-concave function, constrained to the positive orthant with an $l_1$-norm constraint, the maximum is attained when all components are equal. This result leads to the following corollary:
\begin{corollary}
    When the \gls{ms} scheme is applied over \( N \) channels with equal \gls{snr} and a total message size of \( k \) bits, the optimal message allocation \( \bm{k}^* \) is given by
    \begin{align}
        \bm{k}^* = \left( \frac{k}{N}, \frac{k}{N}, \dots, \frac{k}{N} \right).
    \end{align}
\end{corollary}
\section{Numerical Results}
\begin{figure}[htbp]
\centerline{\includegraphics[width=0.9\linewidth]{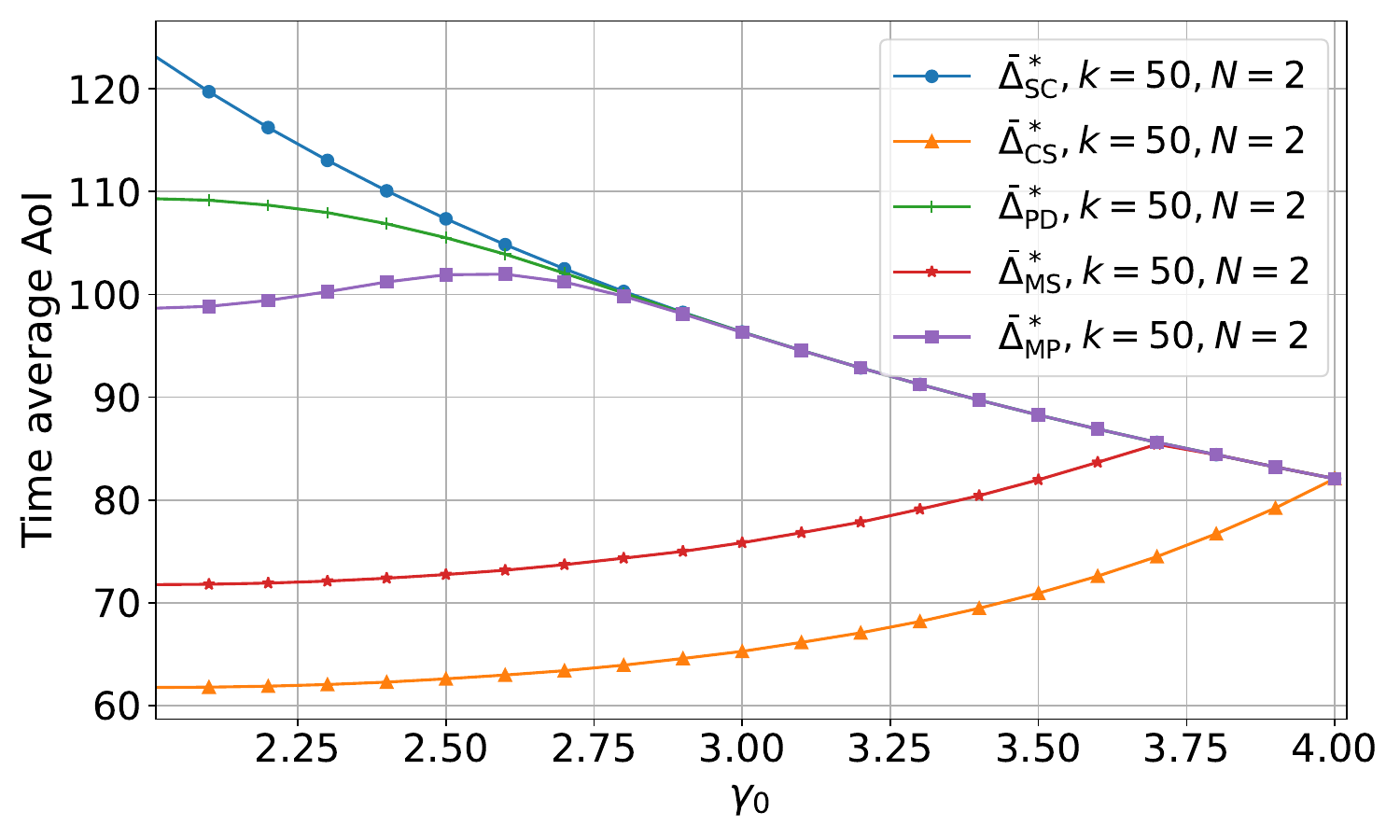}}
\caption{The minimum average AoI achievable by different MC schemes for $N=2$ channels with a sum \gls{snr} of $\gamma_0 + \gamma_1 = 4$. The x-axis represents $\gamma_0$, while $\gamma_1$ is determined as $4-\gamma_0$. }
\label{fig:AoIComparisonDifferentChannels}
\end{figure}

\begin{figure}[htbp]
\centerline{\includegraphics[width=0.9\linewidth]{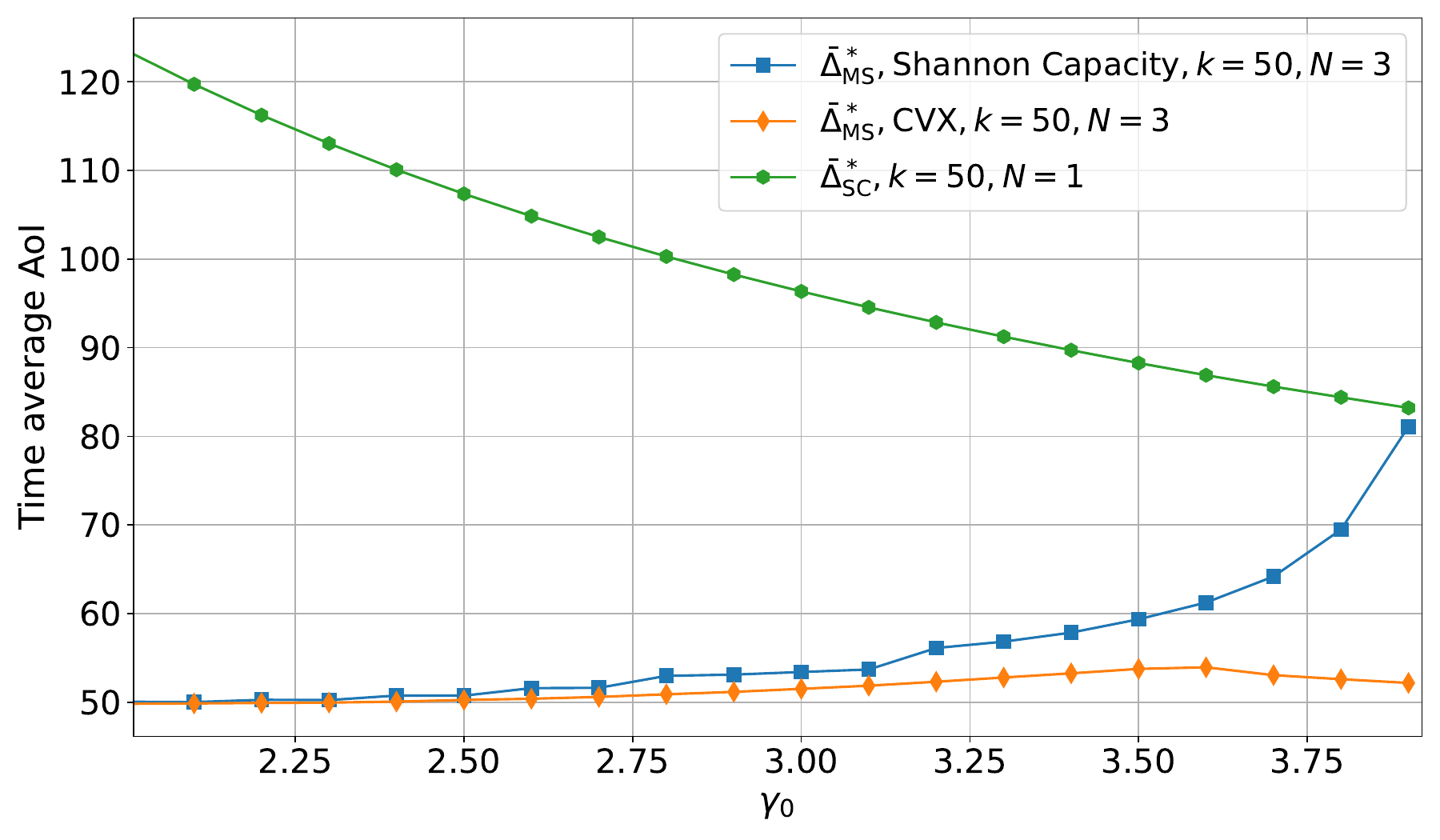}}
\caption{The minimum average \gls{aoi} achievable by different MS schemes for $N=3$ channels with a sum \gls{snr} of $\gamma_0 + \gamma_1 = 4$, where $\gamma_2 = 2.8$. The x-axis represents $\gamma_0$, and $\gamma_1$ is given by $4-\gamma_0$. The average \gls{aoi} of the SC scheme over channel 0 is included for comparison. }
\label{fig:SplitOptimizationNChannels}
\end{figure}

In this section, we numerically compare the average \gls{aoi} achieved by different schemes. In Section \ref{sec:ComparisonEqualSNR}, we previously discussed the performance of the \gls{mc} schemes in the case of $N$ channels with equal \gls{snr}. Here, we extend the analysis to the case where $N$ channels have unequal \gls{snr}. 
We begin by examining how the schemes are influenced by increasingly heterogeneous channels. Specifically, we consider a scenario with $N\!=\!2$ parallel channels, each with unit variance noise and a total power budget of 4. Initially, we set $\gamma_0\! = \!\gamma_1\! =\! 2$ and then gradually increase the \gls{snr} of channel 0 while decreasing that of channel 1, so that $\gamma_0 + \gamma_1\! = \!4$. Fig. \ref{fig:AoIComparisonDifferentChannels} presents the results for this scenario. 
We observe that for the CS scheme, the average \gls{aoi} increases with increasing channel heterogeneity. This behavior is similar to that of the sum capacity of the system. 
%\begin{equation}
%    \frac{1}{2}\log_2(1+\gamma_0) + \frac{1}{2}\log_2(1+\gamma_1).
%\end{equation}
For the \gls{ms} scheme, the average age initially increases as $\gamma_0$ rises but then decreases beyond a certain threshold. This phase transition is driven by the optimal allocation of message bits. Specifically, the transition occurs precisely when the second channel is no longer utilized, i.e., when $k_1 \!=\! 0$. Beyond this point, the system effectively reduces to the SC case with an \gls{snr} of $\gamma_0$.  
In contrast, the average \gls{aoi} of the PD scheme decreases with increasing heterogeneity. For the MP scheme, the average \gls{aoi} initially exhibits a slight increase before eventually decreasing with further channel heterogeneity\footnote{For details on the calculation of the average \gls{aoi} of the \gls{mp} scheme over heterogeneous channels, see Remark 1 in the Appendix.}.
It is important to note that both the PD scheme and the MP scheme, as defined in Section \ref{sec:Systemmodel}, use the same blocklength $n$ across both channels. As a result of this constraint, the error probability of the weaker channel quickly approaches one. In contrast, the MS scheme can balance the error probabilities of the two channels by appropriately distributing the message bits. Similarly, the CS scheme can implicitly adapt to the different channel conditions by appropriately weighting the soft information received from both channels in the decoder. 
Fig. \ref{fig:SplitOptimizationNChannels} compares different methods for allocation of the message bits for the \gls{ms} scheme in a scenario with $N\!=\!3$ channels and unequal \gls{snr} values. The \gls{snr} $\gamma_2 \!=\! 2.8$ is fixed, while the \gls{snr} values of channels 0 and 1 are varied as in Fig. \ref{fig:AoIComparisonDifferentChannels}. The orange curve shows the solution to (\ref{eq:MessageSplitTransformedOptimization}) obtained using CVX. The blue curve represents a baseline scheme where message bits are allocated in proportion to the Shannon capacities of the individual channels relative to the sum channel capacity of all channels together. This scheme performs poorly when the heterogeneity between channels is high. For comparison, the average \gls{aoi} of the \gls{sc} scheme operating over the stronger channel 0 is also included.
\section{Conclusion}
In this paper, we investigate the MC schemes of multiplexing, message splitting, and packet duplication in the context of timely status update systems with short payloads, focusing on the average \gls{aoi} that can be achieved by these schemes. Additionally, we propose a codeword splitting scheme in which each status update is encoded and decoded jointly across all channels. To model transmission errors due to short blocklengths, we employ the second-order normal approximation.
For channels with equal SNR, we analytically establish the ordering of the \gls{mp}, \gls{cs}, and \gls{pd} schemes with respect to the average AoI. Our proposed \gls{cs} scheme achieves the lowest \gls{aoi}, when joint coding is possible. Our numerical evaluations further demonstrate that, depending on the specific parameters, either the MS scheme or the MP scheme can achieve the lowest average \gls{aoi} when joint coding is not possible. For heterogeneous channels, we show how to optimize the distribution of message bits for the \gls{ms} scheme.

\appendices

\section{Proof of Theorem \ref{theo:multiplexingConvexity}}\label{ap:proofconvexityMS}

We start with the $N$ shift parameters $\delta_0, \delta_1,...,\delta_{N-1} < n$, which define the cyclic schedule for the \gls{mp} scheme. These parameters have to fulfill the ordering condition:
\begin{align}
0 = \delta_0 \leq \delta_1 \leq \dots \leq \delta_{N-1} < n.
\end{align}
It is convenient to define the inter-shift (or "waiting") times as
\begin{align}
T_i =
\begin{cases} 
	\delta_{i+1} - \delta_i, & i = 0, \dots, N-2, \\
	n - \delta_{N-1}, & i = N-1.
\end{cases}
\end{align}
Thus, by construction,
\begin{align}
\sum_{i=0}^{N-1} T_i = n.
\end{align}
\begin{figure}[t]
\centerline{\includegraphics[width=0.8\linewidth]{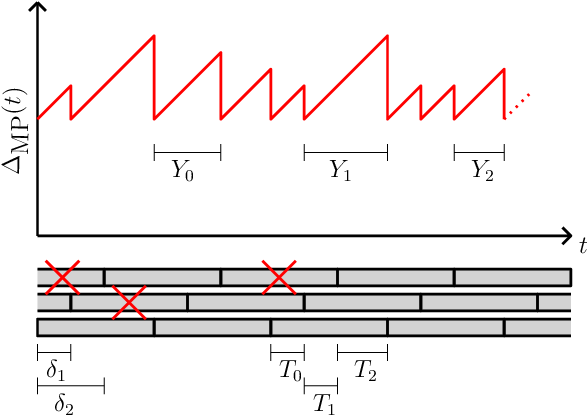}}
\caption{Schematic illustration of the \gls{mp} scheme with \( N = 3 \) channels. The shift parameter \( \delta_0 \) is not shown, as it is defined to be zero by convention. In this example, two status updates are lost on channel 2, and one update is lost on channel 1.}
\label{fig:MultiplexingProofTheorem1}
\end{figure}
For each channel \( i \), let \( Y_i \) be the inter-reception time starting from a successful update on channel \( i \) until the next successful update on any channel. Due to the arbitrary shift parameters, the random variables $Y_i$ are not identically distributed. The shift parameters, waiting times, and inter-reception times $\delta_i, T_i, Y_i$ are illustrated in Fig. \ref{fig:MultiplexingProofTheorem1}.
Note that due to the structure of the cyclic schedule, the random variables $Y_i$ can be defined as follows:
\begin{align}\label{eq:recursive structure Y_i}
    Y_i = T_i + X\cdot Y_{(i+1) \bmod N}, 
\end{align}
where $X$ is a Bernoulli random variable with $p(X=1) = 1-\epsilon$. 
Now consider the area under the age curve starting at a successful transmission over channel $i$ and ending at the next successful transmission on any channel. The expected value of the area under the curve is given by $\frac{1}{2}\mathbb{E}[Y_i^2] + n \mathbb{E}[Y_i]$. Therefore, the expected time average age $\Delta_i$ corresponding to this inter-reception time interval $Y_i$ is given by 
\begin{align}
	\Delta_i = \frac{\mathbb{E}[Y_i^2]}{2\mathbb{E}[Y_i]} + n.
\end{align}
The average \gls{aoi} of the \gls{mp}-scheme can be calculated by weighting the expected time average \gls{aoi} $\Delta_i$ of each inter-reception time $Y_i$ by the expected length of that inter-reception time in relation to the overall time.  Thus, the time average AoI can be written as follows:
\begin{align}\label{eq:BasicMultiplexingAoIExpression}
	\bar{\Delta} & = \sum_{i=0}^{N-1} \frac{(1-\epsilon(n))\mathbb{E}[Y_i]}{\sum_{j=0}^{N-1}(1-\epsilon(n))\mathbb{E}[Y_j]} \Delta_i \\
	& =  \sum_{i=0}^{N-1} \frac{\mathbb{E}[Y_i]}{\sum_{j=0}^{N-1}\mathbb{E}[Y_j]} \Delta_i \\
    & = \frac{1}{2}\sum_{i=0}^{N-1} \frac{ \mathbb{E}[Y_i^2]}{\sum_{j=0}^{N-1}\mathbb{E}[Y_j]} + n. \label{eq:SumOfEYi}
\end{align}
Therefore, we have to calculate the expected inter-reception times and expected squared inter-reception times. 
In the following, we use $\epsilon$ for the error probability $\epsilon(n)$. 

Using (\ref{eq:recursive structure Y_i}), we can formulate a system of $N$ linear equations for the $\mathbb{E}[Y_i]$:
\begin{align}
\mathbb{E}[Y_i] = T_i + \epsilon \mathbb{E}[Y_{(i+1) \bmod N}], \quad i = 0,1,\dots, N-1 .
\end{align}
Note that $ \mathbb{E}[Y_{N\bmod N}] = \mathbb{E}[Y_0]$. 
We can solve this system by unrolling the recursion. For example, starting from \( i = 0 \):
\begin{align}
\mathbb{E}[Y_0] & = T_0 + \epsilon \mathbb{E}[Y_1] \\
& = T_0 + \epsilon \left( T_1 + \epsilon \mathbb{E}[Y_2] \right) \\
 & \vdots \nonumber \\
& = \sum_{j=0}^{N-1} \epsilon^j T_j + \epsilon^N \mathbb{E}[Y_0].
\end{align}
Rearranging gives
\begin{align}
\mathbb{E}[Y_0] (1 - \epsilon^N) = \sum_{j=0}^{N-1} \epsilon^j T_j,
\end{align}
so that
\begin{align}
\mathbb{E}[Y_0] = \frac{\sum_{j=0}^{N-1} \epsilon^j T_j}{1 - \epsilon^N}.
\end{align}
By a similar unrolling (or by cyclic symmetry), the expression for any channel \( i \) is
\begin{align}
\mathbb{E}[Y_i] = \frac{\sum_{j=0}^{N-1} \epsilon^j T_{(i+j) \bmod N}}{1 - \epsilon^N}.\label{eq:YiFirstMoment}
\end{align}
Next, we continue with the expected squared inter-reception times. 
For \(i = 0\), the recurrence is:
\begin{align}
\mathbb{E}[Y_0^2] = T_0^2 + 2\epsilon T_0 \mathbb{E}[Y_1] + \epsilon \mathbb{E}[Y_1^2].
\end{align}
For \(i = 1\), we have:
\begin{align}
\mathbb{E}[Y_1^2] = T_1^2 + 2\epsilon T_1 \mathbb{E}[Y_2] + \epsilon \mathbb{E}[Y_2^2].
\end{align}
Substituting this into the equation for \(\mathbb{E}[Y_0^2]\):
\begin{align}
\mathbb{E}[Y_0^2] & = T_0^2 + 2\epsilon T_0 \mathbb{E}[Y_1] + \epsilon \left( T_1^2 + 2\epsilon T_1 \mathbb{E}[Y_2] + \epsilon \mathbb{E}[Y_2^2] \right) \\
& = T_0^2 + 2\epsilon T_0 \mathbb{E}[Y_1] + \epsilon T_1^2 + 2\epsilon^2 T_1 \mathbb{E}[Y_2] + \epsilon^2 \mathbb{E}[Y_2^2].
\end{align}
Continuing this process for \(N\) steps, we obtain:
\begin{align}
\mathbb{E}[Y_0^2] \!  =\!\!\! \sum_{j=0}^{N-1} \!   \epsilon^j T_j^2 \! + \! 2 \!  \sum_{j=0}^{N-1} \!  \epsilon^{j+1} T_j \mathbb{E}[Y_{(j+1) \bmod N}] + \epsilon^N \mathbb{E}[Y_0^2].
\end{align}
%where indices are taken modulo \(N\).
Rearranging and generalizing this to an arbitrary $i$ gives
\begin{align}
\mathbb{E}[Y_i^2] &= \frac{1}{1 - \epsilon^N} \bigg[ \sum_{j=0}^{N-1} \epsilon^j T_{(i+j) \bmod N}^2  \label{eq:EYiSquared}\\
&\quad + 2 \sum_{j=0}^{N-1} \epsilon^{j+1} T_{(i+j) \bmod N} \mathbb{E}[Y_{(i+j)+1 \bmod N}] \bigg].\nonumber
\end{align}
We now go back to the denominator in (\ref{eq:SumOfEYi}). By the geometric series, we have
\begin{align}
	\sum_{i=0}^{N-1} \mathbb{E}[Y_i] & = \sum_{i=0}^{N-1} \frac{1}{1-\epsilon^N} \sum_{j=0}^{N-1} \epsilon^j T_{(i+j)\bmod N} \\
	& = \frac{1}{1-\epsilon^N} \sum_{j=0}^{N-1} \epsilon^j \sum_{i=0}^{N-1} T_i \\
	& =  \frac{1}{1-\epsilon^N} \sum_{j=0}^{N-1} \epsilon^j n \\
	& =  \frac{1}{1-\epsilon^N} \frac{1-\epsilon^N}{1-\epsilon} n \\
	& = \frac{n}{1-\epsilon}.
\end{align}
Inserting this in (\ref{eq:SumOfEYi}), the average AoI can be written as
\begin{align}
	\bar{\Delta} = \frac{1-\epsilon}{2n} \sum_{j=0}^{N-1} \mathbb{E}[Y_j^2] + n .\label{eq:aoiwithSquaredYi}
\end{align}
Inserting (\ref{eq:YiFirstMoment}) into (\ref{eq:EYiSquared}) and (\ref{eq:EYiSquared}) into (\ref{eq:aoiwithSquaredYi}), we obtain the following expression for time average AoI:
\begin{align}
	\bar{\Delta} &= \frac{1-\epsilon}{2n(1-\epsilon^N)} \Bigg\{ 
	\sum_{i=0}^{N-1} \sum_{j=0}^{N-1} \epsilon^j\,T_{(i+j) \bmod N}^2 
	\Bigg\} \notag \\
	&\quad + \frac{1-\epsilon}{2n(1-\epsilon^N)}  \frac{2}{1-\epsilon^N} \Bigg\{ 
	\sum_{i=0}^{N-1} \sum_{j=0}^{N-1} \epsilon^{j+1}\,T_{(i+j) \bmod N} \notag \\
	&\quad \times \left(\sum_{k=0}^{N-1}\epsilon^k\,T_{(i+j+1+k) \bmod N}\right)
	\Bigg\} + n \label{eq:AoIMultiplexingDetailedNChannels} \\
    & = \frac{1-\epsilon}{2n(1-\epsilon^N)} \sum_{i=0}^{N-1} \Bigg\{
	\frac{1-\epsilon^N}{1-\epsilon} T_i^2 \notag \\
	&\quad + \frac{2}{1-\epsilon^N} \sum_{j=0}^{N-1} \epsilon^{j+1}\,T_{(i+j) \bmod N} \notag \\
	&\quad \times \left(\sum_{k=0}^{N-1} \epsilon^k\,T_{(i+j+1+k) \bmod N}\right)
	\Bigg\} + n,\label{eq:AoIMultiplexingDetailedNChannelsSimplified}
\end{align}
where we have applied the geometric series on the first double sum in (\ref{eq:AoIMultiplexingDetailedNChannels}). 
\begin{remark}\label{rem:MPUnequal Channels}
The expression in (\ref{eq:BasicMultiplexingAoIExpression}) remains valid when the error probabilities $\epsilon_i$ are unequal. In this case, the corresponding systems of linear equations for the expected inter-reception times $\mathbb{E}[Y_i]$ and their second moments $\mathbb{E}[Y_i^2]$ still hold, provided that each occurrence of $\epsilon$ is replaced by the specific $\epsilon_i$. These equations can be solved numerically.  Note that the expressions for the error probabilities in (\ref{eq:BasicMultiplexingAoIExpression}) have to be replaced as well. Substituting the numerically obtained values of $\mathbb{E}[Y_i]$ and $\mathbb{E}[Y_i^2]$ into (\ref{eq:BasicMultiplexingAoIExpression}) yields the average \gls{aoi} for the \gls{mp} scheme operating over heterogeneous channels.
\end{remark}
We now want to show that this function is convex on $\mathcal{R}_+^N$ with $l_1$ norm constraint $\sum_{i=0}^{N-1}T_i = n$. Therefore, we take a closer look at the second part of (\ref{eq:AoIMultiplexingDetailedNChannelsSimplified}):
\begin{align}\label{eq:AoIExpressionsNChannelsSecondPart}
	& \frac{2}{1-\epsilon^N} \!\! \sum_{i=0}^{N-1} \sum_{j=0}^{N-1}\!\!\epsilon^{j+1}\,T_{(i+j) \bmod N}\!\left(\sum_{k=0}^{N-1}\!\epsilon^k\,T_{(i+j+1+k) \bmod N}\!\right) \\
	& = \frac{2}{1-\epsilon^N} \sum_{j=0}^{N-1} \epsilon^{j+1}\sum_{r=0}^{N-1} T_r \left(\sum_{k=0}^{N-1}\epsilon^k\,T_{(r+1+k) \bmod N}\right) \\
	& = \frac{2\epsilon}{1-\epsilon} \sum_{r=0}^{N-1} T_r \left(\sum_{k=0}^{N-1}\epsilon^k\,T_{(r+1+k) \bmod N}\right).
\end{align}
This function is cyclic invariant in the variables $T_0, T_1, ..,T_{N-1}$. 
To determine convexity, (\ref{eq:AoIMultiplexingDetailedNChannelsSimplified}) has to be expressed as a quadratic form and the corresponding matrix has to be checked for \gls{psd}. Let $\bm{T} = (T_0, T_1,..,T_{N-1})^T$ be the vector containing the waiting times. 
We can express (\ref{eq:AoIExpressionsNChannelsSecondPart}) as $\mathbf{T}^T \mathbf{Q} \mathbf{T}$ with $\mathbf{Q} = \frac{1}{2}(\mathbf{A} + \mathbf{A}^T)$ \cite[Example 4.0.2]{horn13}.
The elements of the matrix $\mathbf{A}$ are defined as
\begin{align}
	a_{i,j} & = \frac{2\epsilon}{1-\epsilon} \epsilon^{j-i-1 \text{ mod } N}
\end{align}
The matrices $\mathbf{A}$ and $\mathbf{Q}$ are given as
%\begin{align}
	%\mathbf{A} = \frac{2\epsilon}{1-\epsilon}
	%\begin{pmatrix}
		%\epsilon^{N-1} & 1           & \epsilon     & \epsilon^2   & \cdots & \epsilon^{N-2} \\
		%\epsilon^{N-2} & \epsilon^{N-1} & 1         & \epsilon     & \cdots & \epsilon^{N-3} \\
		%\epsilon^{N-3} & \epsilon^{N-2} & \epsilon^{N-1} & 1     & \cdots & \epsilon^{N-4} \\
		%\vdots         & \vdots      & \vdots       & \vdots       & \ddots & \vdots \\
		%1              & \epsilon   & \epsilon^2 & \epsilon^3  & \cdots & \epsilon^{N-1}
	%\end{pmatrix}\,.
%\end{align}
\begin{align}
	\mathbf{A} = \frac{2\epsilon}{1-\epsilon}
	\begin{pmatrix}
		\epsilon^{N-1} & 1           & \epsilon     & \cdots & \epsilon^{N-2} \\
		\epsilon^{N-2} & \epsilon^{N-1} & 1         & \cdots & \epsilon^{N-3} \\
		\epsilon^{N-3} & \epsilon^{N-2} & \epsilon^{N-1} & \cdots & \epsilon^{N-4} \\
		\vdots         & \vdots      & \vdots       & \ddots & \vdots \\
		1              & \epsilon   & \epsilon^2 & \cdots & \epsilon^{N-1}
	\end{pmatrix}\,.
\end{align}
and
%\begin{align}
	%& \mathbf{Q} =
	%\frac{1}{2}\Bigl(\mathbf{A}+\mathbf{A}^T\Bigr) \\
	%& = \frac{1}{1-\epsilon}
	%\begin{pmatrix}
		%2\epsilon^{N} & \epsilon+\epsilon^{N-1} & \cdots & \epsilon+\epsilon^{N-1}\\[1mm]
		%\epsilon+\epsilon^{N-1} & 2\epsilon^{N} & \cdots & \epsilon^2+\epsilon^{N-2}\\[1mm]
		%\epsilon^2+\epsilon^{N-2} & \epsilon+\epsilon^{N-1} & \cdots & \vdots\\[1mm]
		%\vdots & \vdots & \ddots & \vdots\\[1mm]
	%\epsilon+\epsilon^{N-1} & \cdots & \cdots & 2\epsilon^{N}
	%\end{pmatrix}\,.
%\end{align}
\begin{align}
	& \mathbf{Q} =
	\frac{1}{2}\Bigl(\mathbf{A}+\mathbf{A}^T\Bigr) \\
	& = \frac{1}{1-\epsilon}
	\begin{pmatrix}
		2\epsilon^{N} & \epsilon+\epsilon^{N-1} & \cdots & \epsilon+\epsilon^{N-1}\\[1mm]
		\epsilon+\epsilon^{N-1} & 2\epsilon^{N} & \cdots & \epsilon^2+\epsilon^{N-2}\\[1mm]
		\vdots & \vdots & \ddots & \vdots\\[1mm]
	\epsilon+\epsilon^{N-1} & \cdots & \cdots & 2\epsilon^{N}
	\end{pmatrix}\,.
\end{align}
Now the contribution of $\sum_{i=0}^{N-1}$ $\frac{1-\epsilon^N}{1-\epsilon} T_i^2$, which is the first part in the large brackets of (\ref{eq:AoIMultiplexingDetailedNChannelsSimplified}), has to be added to the quadratic form. Therefore the term $\frac{1-\epsilon^N}{1-\epsilon}$ is added to the main diagonal and overall $\bar{\Delta}$ can be written as follows:
\begin{align}
	\bar{\Delta} = \frac{1-\epsilon}{2n(1-\epsilon^N)} \bm{T}^T \bm{M} \bm{T} + n,
\end{align}
with the real symmetric matrix $\bm{M}$ given as 
%\begin{align}
	%\mathbf{M} = \\
    %& \frac{1}{1-\epsilon}
	%\begin{pmatrix}
		%1+\epsilon^N & \epsilon+\epsilon^{N-1} & \cdots & \epsilon+\epsilon^{N-1}\\[1mm]
		%\epsilon+\epsilon^{N-1} & 1+\epsilon^N & \cdots & \epsilon^2+\epsilon^{N-2}\\[1mm]
		%\epsilon^2+\epsilon^{N-2} & \epsilon+\epsilon^{N-1} & \cdots & \vdots\\[1mm]
		%\vdots & \vdots & \ddots & \vdots\\[1mm]
		%\epsilon+\epsilon^{N-1} & \epsilon^2+\epsilon^{N-2}  & \cdots & 1+\epsilon^N
	%\end{pmatrix}\,. \nonumber
%\end{align}
\begin{align}
	\mathbf{M} = \\
    & \frac{1}{1-\epsilon}
	\begin{pmatrix}
		1+\epsilon^N & \epsilon+\epsilon^{N-1} & \cdots & \epsilon+\epsilon^{N-1}\\[1mm]
		\epsilon+\epsilon^{N-1} & 1+\epsilon^N & \cdots & \epsilon^2+\epsilon^{N-2}\\[1mm]
		\vdots & \vdots & \ddots & \vdots\\[1mm]
		\epsilon+\epsilon^{N-1} & \epsilon^2+\epsilon^{N-2}  & \cdots & 1+\epsilon^N
	\end{pmatrix}\,. \nonumber
\end{align}
The Hessian of a quadratic form with a symmetric matrix $\bm{M}$ is given by $2\bm{M}$ \cite[page 644]{boydConvexOpt2006}.  The factor 2 does not change if the matrix is PSD. We can therefore investigate if the matrix $\bm{M}$ is PSD. Note that the matrix is circulant. The eigenvalues $\lambda_0, \lambda_1,..,\lambda_{N-1}$ of a circulant matrix are given by the following formula \cite[Theorem 3.1]{GrayToeplitz2005}:
\begin{align}
	\lambda_k = m_0 + m_1\omega^{k}_N + m_2 \omega^{2k}_N + ... +m_{N-1} \omega^{(N-1)k}_N, 	
\end{align}
where the $m_j$ are the entries in the first row of the matrix and $\omega_N = e^{-\frac{2\pi i}{N}}$. For the matrix $\bm{M}$, the coefficients are given by
\begin{align}
	m_j = \frac{\epsilon^j + \epsilon^{N-j}}{1-\epsilon}. 
\end{align}
Thus we have
\begin{align}
	\lambda_k & = \frac{1}{1-\epsilon} \sum_{j=0}^{N-1} \left(  \epsilon^j + \epsilon^{N-j} \right) \omega^{jk}_N \\
	& = \frac{1}{1-\epsilon}\left[ S_1(k) + S_2(k)  \right].
\end{align}
We start with $S_1(k)$. Using the geometric series, we have
\begin{align}
	S_1(k) & = \sum_{j=0}^{N-1} (\epsilon \omega^k_N)^j \\
	& = \frac{1-(\epsilon \omega^k_N)^N}{1 -\epsilon \omega^k_N} \\
	& = \frac{1-\epsilon^N}{1 -\epsilon \omega^k_N}. 
\end{align}
For the second sum, we have
\begin{align}
	S_2(k) & = \sum_{j=0}^{N-1} \epsilon^{N-j} \omega^{jk}_N \\
	& = \sum_{l=1}^{N} \epsilon^l \omega^{(N-l)k}_N \\
	& = \sum_{l=1}^{N} \epsilon^l \omega^{-lk}_N \\
	& = \epsilon \omega^{-k}_N \sum_{l=0}^{N-1} (\epsilon\omega^{-k}_N)^l \\
	& =  \epsilon \omega^{-k}_N \frac{1-(\epsilon \omega^{-k}_N)^N}{1- \epsilon \omega^{-k}_N} \\
	& = \epsilon \omega^{-k}_N \frac{1-\epsilon^N}{1- \epsilon \omega^{-k}_N}.
\end{align}
Combining both sums results in the following expression for the eigenvalues of $\bm{M}$:
\begin{align}
	\lambda_k & = \frac{1-\epsilon^N}{1-\epsilon} \left[ \frac{1}{1-\epsilon \omega^{k}_N} + \frac{\epsilon \omega^{-k}_N}{1-\epsilon \omega^{-k}_N}  \right] \\
	& = \frac{1-\epsilon^N}{1-\epsilon} \left[ \frac{1-\epsilon \omega^{-k}_N + \epsilon \omega^{-k}_N(1-\epsilon \omega^{k}_N)}{(1-\epsilon \omega^{k}_N) (1-\epsilon \omega^{-k}_N)}  \right] \\
	& = \frac{1-\epsilon^N}{1-\epsilon} \left[ \frac{1-\epsilon^2}{1-2\epsilon \cos(2\pi k /N) + \epsilon^2}  \right]. 
\end{align}
From this, we can conclude that all the eigenvalues are greater than 0 if $0 < \epsilon < 1$, ensuring the strict convexity of the function. Consequently, the function has a unique global minimum. Furthermore, since the function is invariant under cyclic shifts of the variables, it must attain its global minimum at the symmetric point $T_0 = T_1 = .. =T_{N-1} = n/N$.

\bibliographystyle{IEEEtran}
\IEEEtriggeratref{9}
\bibliography{IEEEabrv,AoIMCGlobeCom}

\end{document}